# Hydrogen permeation on defected $\alpha$-Al$_2$O$_3$ surfaces: DFT calculations

**Xueyan Wang[1], Man Jiang[1] and Zhangcan Yang[1]**

[1] Department of Nuclear Engineering and Technology, School of Energy and Power Engineering, Huazhong University of Science and Technology, Wuhan 430074, People's Republic of China

E-mail: yang_zhangcan@hust.edu.cn

**Abstract**

One of the key challenges to realize controlled fusion energy is tritium self-sufficiency. The application of hydrogen permeation barrier (HPB) is considered to be necessary for tritium self-sufficiency. $\alpha$-Al$_2$O$_3$ is currently a candidate material for HPB. However, a crucial issue for $\alpha$-Al$_2$O$_3$ is that its permeability reduction factor (PRF) will dramatically drop after ion or neutron irradiations. At present, little is known about the relevant mechanism. In order to shed light on this issue, the kinetics and energetic changes of hydrogen on defected $\alpha$-Al$_2$O$_3$ surfaces in comparison with perfect $\alpha$-Al$_2$O$_3$ surfaces were studied by density functional theory. For perfect $\alpha$-Al$_2$O$_3$ surfaces, the results show that the barrier for hydrogen migration from the outermost layer into the subsurface layer is the highest, making this migration step to be a rate limiting process. In contrast, surface point defects dramatically reduce this maximum barrier. Consequently, hydrogen can preferentially permeate into the interior of the material through surface defects. The findings can help explain the possible mechanism of significant decrease of PRF under radiation.

## 1. Introduction

Preventing the loss of tritium to coolant and structural materials is crucial to attain tritium self-sufficiency for D-T nuclear fusion power plants [1]. One effective method is to cover structural materials with a thin coating layer acted as a hydrogen permeation barrier (HPB). Various ceramics including $\alpha$-Al$_2$O$_3$, $\alpha$-Cr$_2$O$_3$ [2, 3], Er$_2$O$_3$ [4, 5], TiO$_2$ [6], TiN, TiC [7] and SiC [8] have been found to be good barriers. Among them, $\alpha$-Al$_2$O$_3$ has attracted more and more attention due to its high permeability reduction factor (PRF) [9-13] at working temperatures, high melting point and outstanding mechanical properties.

Although $\alpha$-Al$_2$O$_3$ has shown extremely high PRF in the laboratory, its PRF is 3 to 4 orders of magnitude lower when used in a radiation environment [14, 15]. Understanding how radiation affects the performance of permeation barriers is critical for the successful use of HPB. However, the underlying mechanism accounting for radiation-induced PRF decrease is still unclear. In particular, there lacks of microscopic insight into the interaction of hydrogen isotopes with irradiation-damaged $\alpha$-Al$_2$O$_3$.

In order to understand the effects of radiation on hydrogen permeation, we need to first examine how perfect $\alpha$-Al$_2$O$_3$ layer resists hydrogen permeation. For hydrogen isotopes to permeate through an $\alpha$-Al$_2$O$_3$ coating layer, hydrogen molecules must adsorb on the surfaces, dissociate into atoms, dissolve into the layer, diffuse through the layer, and then diffuse to the base material [16]. The rates of these processes can be distinct, varying a few orders of magnitude. Zhang et al.[17] studied the interaction between H atom and a perfect (0001) surface of $\alpha$-Al$_2$O$_3$ through DFT method. They calculated the potential pathways of H diffusion and predicted that H migrating from the outermost surface layer to the subsurface is the rate-controlling process due to the highest energy barrier for this process.

It is known that radiation can induce various kinds of defects in crystal materials including ceramics. Schottky defects in full charged state are found to be the most common and stable defects when α-$Al_2O_3$ surfaces are irradiated. Devanathan *et al* [18] and Matzke *et al* [19] showed that point defects can be easily produced in ceramic materials at low radiation dose or low energy heavy ion impact. As the radiation dose or the energy of the incident particles increases, defect clusters or complex defect structures are produced. Besides, Carrasco et al. [20], through first-principles calculations, indicated that the formation energy of point defects on a surface is significantly lower than the same defects in interior environment, implying that point defects on a surface are more stable than when they are inside a material. Floro et al. [21] observed the migration of internal defects to surface in experiments, in consistent with the DFT calculations by Carrasco et al. [20]. As for the types of the point defects, Matsunaga et al. [22] found through first-principles calculations that in α-$Al_2O_3$, the formation energy of Schottky defects is significantly lower than that of Frenkel defects, implying Schottky defects are dominant in pure α-$Al_2O_3$. Moreover, they found that a defect in full charged state is most stable in the bulk environment.

Point defects have been found to strongly interact with H, affecting the trapping and diffusion mechanisms of H. As shown in a study of lithium oxide [23], H migration is mediated by vacancy point defects. Zhang et al [24] studied H interactions with intrinsic point defects in bulk α-$Al_2O_3$. According to their calculations, oxygen vacancies are strong traps for H, which decreases the H mobility and eventually the H permeation at non-equilibrium conditions such as radiations. This finding seems to be contradictory with experimental observation of increase of permeation under radiation. A possible explanation is speculated here. Zhang's calculations are for bulk α-$Al_2O_3$. As mentioned before, the process of H diffusion from the surface to subsurface is the rate-controlling process. Surface defects produced by radiation may significantly lower the migration barrier for H dissolution into the material. As a result, the decrease of H diffusion in bulk may be fully compensated by the increase of H diffusion from the surface to subsurface, resulting a net effect of increase in H permeation. Besides, Zhang et al [24] only calculated the interaction of single H with point defects. It is unclear how many hydrogen can fill into a trap and how the migration barrier changes when more than one hydrogen is trapped in a defect.

In order to verify this hypothesis, we will investigate the interactions between H and typical point defects in α-$Al_2O_3$ surface using DFT calculations. The potential pathways of H diffusion in perfect α-$Al_2O_3$ surfaces and defected surfaces will be calculated and compared. Our calculations will provide a useful reference for further exploration of the mechanism of decreasing PRF under radiation.

## 2. Computational details

### 2.1 Method and parameter settings

All present calculations were performed with DFT plane wave method [25] utilizing quantum chemical software Vienna Ab initio Simulation Package (VASP) [26, 27]. The pseudopotential of the elements were from the generalized gradient approximation (GGA) using the PW91 functional [28, 29]. The interaction between solid ions and valence shell electrons was described by the projector augmented wave (PAW) method [27, 30]. The Brillouin zone was sampled by gamma points due to trigonal structure of α-$Al_2O_3$ [31]. Calculation of the lattice structure optimization was performed using 3×3×2 K-point grid, and all calculations about surface were performed using 3×3×1 K-point grid. These grids above were found to be sufficiently accurate for sampling Brillouin zone of the supercells. The cut-off energy was set to be 520 eV constantly [32]. The Gaussian smearing method was used to achieve better convergence, with the Gaussian temperature spread set to be 0.2 eV. The effects of spin polarization were considered. All atoms were allowed to relax during structural optimization. The atomic coordinates were relaxed until Hellmann-Feynman forces [33, 34] acting on each atom were reduced to less than 0.02 eV/Å. The climbing image nudged elastic band (CINEB) method [35] was used to find the minimum energy path and transition state in H atoms' migration. For each pivotal site we found (including stable and transition sites), it was also verified by performing a frequency calculation.

### 2.2 Model configuration

#### 2.2.1 Lattice optimization

Through the structural energy minimization of the lattice in bulk, the lattice constant is a = b = 4.802 Å, c = 13.101 Å, which agrees well with other first-principles calculation results [17, 36, 37], and experimental value [38]. There are totally two kinds of Al-O bonds inside α-$Al_2O_3$, with the length of 1.989 Å for the longer type and 1.871 Å for the shorter type. The total energy of α-$Al_2O_3$ is $3.644×10^3$ KJ/mol.

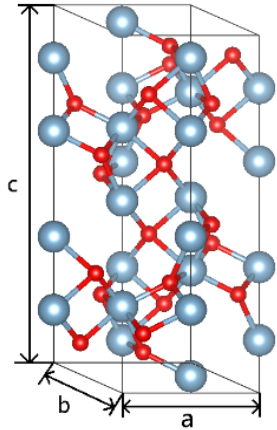

**Fig. 1-α-Al$_2$O$_3$ lattice**

**Table 1-**

| parameter | our | DFT-1 [17] | DFT-2 [36] | DFT-3 [37] | experimental [38] |
|---|---|---|---|---|---|
| a&b/Å | 4.802 | 4.806 | 4.834 | 4.794 | 4.759 |
| c/Å | 13.101 | 13.133 | 13.164 | 13.100 | 12.990 |

Due to the periodic boundary conditions of the supercell, the size of the supercell must be large enough to ensure that the H atoms and defect inside the supercell enjoy little interaction with H atoms and defects in adjacent supercells [20]. The bottom edge size of the supercell was set to be 9.605 Å = 2a = 2b in the subsequent calculations.

*2.2.2 Perfect surface model*

For actual material's surface, in most cases, is the most stable type, which has already been confirmed by some crystal growth experiments [39, 40]. Especially for α-Al$_2$O$_3$, the coordination relationship of the atoms and electronic structure on the surface are quite different from internal bulk environment [41]. Therefore, the most stable surface should be adopted in the slab model. In all kinds of α-Al$_2$O$_3$ crystal surfaces, the surface in (0001) direction is the most stable compared to (1-102) and (11-20) [42-44]. The α-Al$_2$O$_3$ (0001) surface may be considered as a paradigm of a complex surface and, consequently, has been the subject of many experimental and theoretical studies [20], the (0001) surface can be more specifically divided into 3 types: a single Al layer terminating surface (a), a double Al layer terminating surface (b), and an O layer terminating surface (c), as shown in Figure 2.

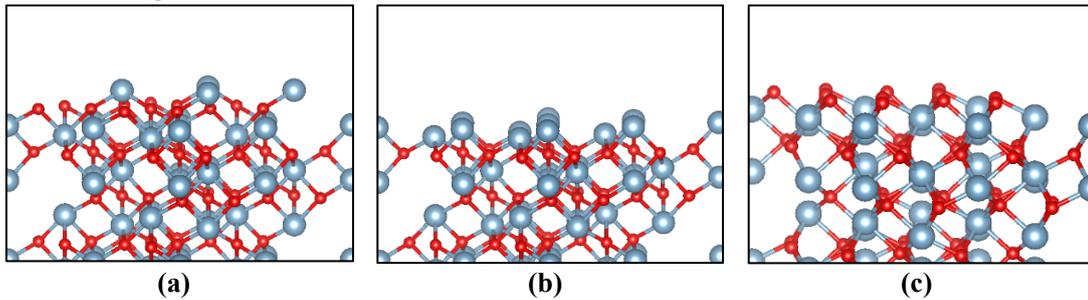

**(a)**      **(b)**      **(c)**

**Fig. 2-three different kinds of surface of α-Al$_2$O$_3$ in (0001) direction**

By relaxation, the surface energies of these three surfaces are 1.58 J/m$^2$, 6.18 J/m$^2$ and 6.07 J/m$^2$, respectively, which are in good agreement with the known calculation results [17, 37, 42, 45]. It can be seen that the most stable surface is a single Al layer terminating surface.

Surface structures were modeled as periodic slabs, thus it's indispensable to let the vacuum layer wide enough to ensure the interaction between adjacent slabs is sufficiently small which can be omitted [46]. For the surface model obtained, a pair of control experiments was set up, set the thickness of the vacuum layer to be 7.5 Å and 15 Å respectively (15 Å is wide enough to make the slab-to-slab effect negligible). Through comparison, we find that the relative difference in total energy between the two systems is less than $2.2\times10^{-5}$, which is indicative of that it is feasible to set a vacuum layer with width of 7.5 Å.

Setting up the thickness of the slab is a crucial issue. If too thin, the two surfaces of the slab will be too close and cause interaction effect, which will make the calculation results deviate from reality. To avoid this, a set of calculations was performed to find a value large enough for thickness. As shown in Figure 3, the H atom migrates from a stable adsorption site on the outermost layer to the interior of the slab, sequentially records the binding energy of each stable site along the path. As can be seen from the graph, before the H atom reaching site 6, the binding energy will change with sites. Except site 1, site 4 has the lowest energy and is the most stable. As the H atom moves deeper past site 6, its formation energy gradually approach to a constant value, approximately 4.02 eV, while $E_{rel}$ gradually approach 2.60 eV. That is, in this range, it can be considered that the environment in which the H atom migrates is bulk environment. Mark the depth of site 6 as d, as demonstrated by leads in Figure 3 below.

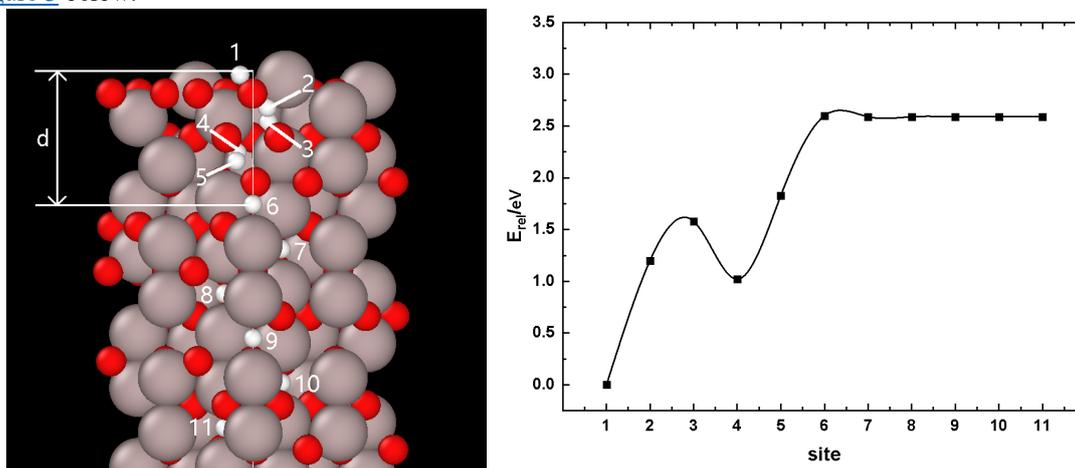

**Fig. 3- $E_{rel}$ of H atom corresponding to serial stable sites, $E_{rel}$ represents the change value of each site's energy along the path relative to site 1.**

Our results here ( from site 6 to site 11 in Figure 3 ) show that H atom has one and only one kind of stable site in bulk environment, and does not form a bond with any surrounding atoms, meanwhile, the migration barrier between two adjacent stable sites is approximately 1.08 eV. We believe that there should be a transitional region which approximates bulk environment to avoid interaction effects between the two opposite surface regions, which requires the slab model to be thick enough. The thickness of the slab should not be thinner than 2d, which is exactly close to the height of a lattice (13.1 Å). Zhang et al. [17] showed that through the α-$Al_2O_3$ slab model they built, H atom invariably has tendency to form bonds with other atoms throughout the migration process. The model they used contained only four layers of O atoms, perhaps the slab's span in the direction perpendicular to the surface is not long enough, i.e. the distance may be too short between the two opposite surfaces of the slab, accordingly, their H atom may has entered the opposite surface region before leaving the original surface region, thus the calculation results of H migration may be somewhat different from the actual situation.

*2.2.3 Analysis of types of defected surface model*

According to the analysis in introduction, since the barrier from the outermost layer to the subsurface is the maximum value throughout the entire permeation path, the outermost layer of atoms on the surface play an important role in preventing H atom from migrating into the slab. Therefore, we discuss the effect on H atom's permeation when there is a point defect in the outermost layer of the slab material.

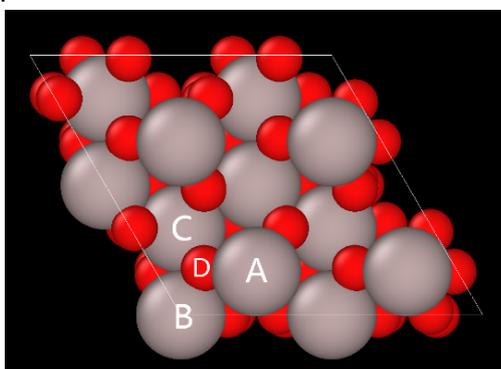

**Fig. 4- Four types of surface point defects, top view**

Analyzing the types of ionic point defects which may exist on the surface is firstly need to be done. Observing the top view of the surface of slab, it can be found that there are three kinds of Al atoms in the outermost layer, as shown by A B C in Figure 4, and only one kind of O atoms, as shown by D. Considering that the most stable Schottky defects are in case of fully charged, the model we built is that the A/B/C/ position on the outermost layer respectively lacked an $Al^{3+}$ and the D position lacked an $O^{2-}$, by setting the number of valence electrons ensures that ion defects are charged after iterative convergence.

## 3. Results and discussion

### 3.1 H atoms permeate from perfect α-$Al_2O_3$ (0001) surface into the slab

### 3.1.1 Energy curve

The surface model from 2.2.2 was used. First, the H atom stays at a stable adsorption site (site 1) on the outermost layer, forms a H-O bond with an O atom. We envisage a migration path of H atom from the outermost layer into the interior of the slab, i.e.

$$1 \to TS12 \to 2 \to TS23 \to 3 \to TS34 \to 4 \to TS45 \to 5 \to TS56 \to 6 \to \text{bulk environment.}$$

As shown in Figure 5 below. According to the analysis in 2.2, when the H atom migrates past site 6 and continues to migrate deeper into the slab, it can be considered to enter the bulk environment.

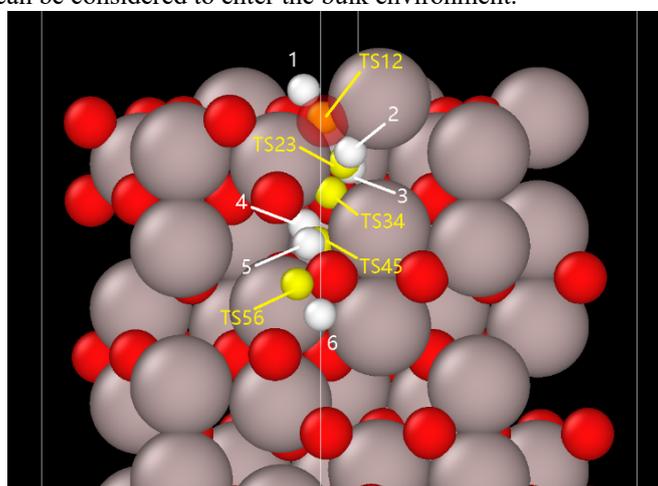

**Fig. 5- Migration path through perfect surface**
**(1) For the convenience of vision, individual atoms are moved slightly, meanwhile, a few Al atoms and O atoms are hidden or translucent to make the path easier to observe.**
**(2) The small spheres with serial numbers are H atoms, where white indicates a stable site and yellow indicates a transition site. (Subsequent sections are the same as here)**

The energy curve and relative energy $E_{rel}$ of serial sites along the path are demonstrated in Figure 6 and Table 2 below.

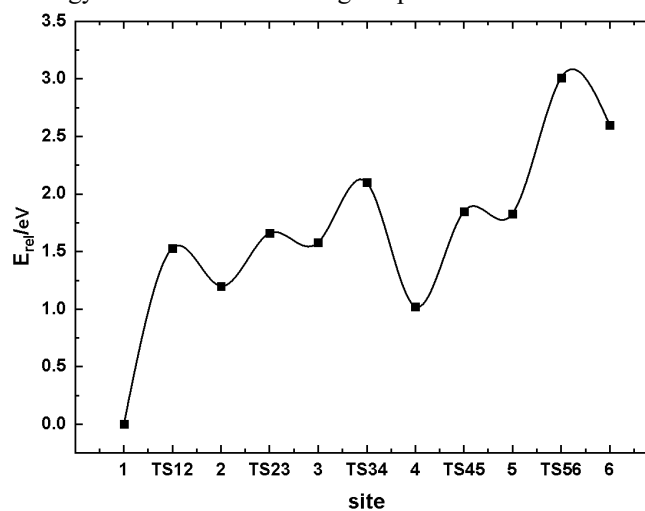

**Fig. 5- Energy curve along the migration path through perfect surface**

**Table 2- Relative energy $E_{rel}$ corresponding to serial sites**

| site | 1 | TS12 | 2 | TS23 | 3 | TS34 | 4 | TS45 | 5 | TS56 | 6 |
|---|---|---|---|---|---|---|---|---|---|---|---|
| $E_{rel}$ /eV | 0 | 1.53 | 1.20 | 1.66 | 1.58 | 2.10 | 1.02 | 1.85 | 1.83 | 3.01 | 2.60 |

Before entering the bulk environment, each step of H atom's migration can be divided into two types: (1) H atom forms a H-O bond with an O atom, and rotates around the O atom to reorient the bond; (2) The bond between H and O atoms is broken and then H atom jumps to another O atom and forms a new H-O bond with it.

Site 1 is a stable adsorption site for the H atom, where H and O atoms form a H-O bond with the length of 1.26 Å. Next the H atom remains bonded and rotated on this O atom for reorientation, migrates to site 2. This step requires significant energy consumption, with the energy barrier of 1.53 eV, which is the highest barrier on the entire path, thus confirmed that the outermost layer of atoms play an important role in preventing H from permeating into the material. Then the H atom migrates along 2→TS23→3→TS34→4, overcomes two not too high barriers, 0.46 eV and 0.52 eV, at sites 2 and 3, reaching site 4. There is a deep well at site 4, which is the site with lowest energy and the highest stability except site 1 throughout the path, the H atom migrates along 4→TS45→5→TS56→6 afterwards, overcomes two higher barriers, 0.84 eV and 1.18 eV, at sites 4 and 5, reaching site 6. The formation energy of H atoms at site 6 can be compared with the value in bulk environment, with a difference of no more than 0.03 eV. Therefore, it can be considered that the diffusion from site 6 to the deeper portion of the slab is in bulk environment.

*3.1.2 Analysis of the causes of a deep well for H atom at site 4*

Generally for crystals, the abrupt termination in the direction perpendicular to the surface of the material causes a dramatic change in the coordination relationship of the atoms in the region near surface. The atoms on the outermost layer can only interact with the secondary surface layer atoms, which makes such interaction stronger than that between the two adjacent layers in bulk environment, thus the distance between outermost layer and secondary surface layer is shortened. Further, the effect of the subsurface layer and the deeper atoms is weakened, thus the distance is increased.

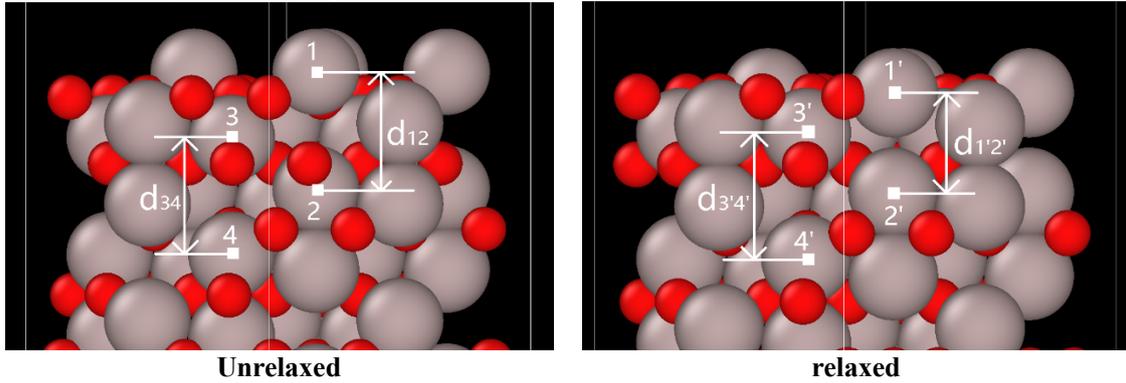

**Unrelaxed**  **relaxed**
**Fig. 6-α-Al₂O₃ (0001) surface structure before and after relaxation**

The structure of the slab model used in this study before and after relaxation are shown in the two parts of Figure 7. 1, 2, 3 and 4 in the left part represent four Al atoms, respectively. After relaxation, they correspond to the 1', 2', 3' and 4' atoms in the right part, $d_{12}$, $d_{1'2'}$, $d_{34}$ and $d_{3'4'}$ respectively represent the distance between Al atoms marked by leads. The slab is fully relaxed and the coordinates of these atoms are recorded. It is found that the distance between 1 and 2 is reduced, $d_{1'2'}-d_{12} = -0.7$ Å, the interaction is enhanced; the distance between 3 and 4 is increased, $d_{3'4'}-d_{34} = 0.2$ Å, the interaction is weakened. This result is consistent with experimental observations [42].

The weakly interacting region between Al atoms 3 and 4 produces a deep well for H atom, such that H atom reaches an evident low energy site (site 4) after entering the subsurface and is thus easily retained at this site. This phenomenon is confirmed with the above analysis mutually.

*3.2 H atoms permeate into α-Al₂O₃ slab from surface with point defects*

*3.2.1 An Al3+ lacked at position A on the outermost layer*

The diffusion path of H atom is:
1→TS12→2→TS23→3→TS34→4→TS45→5→bulk environment.

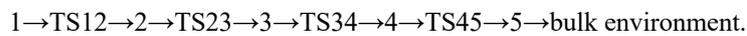

As demonstrated in Figure 8 below.

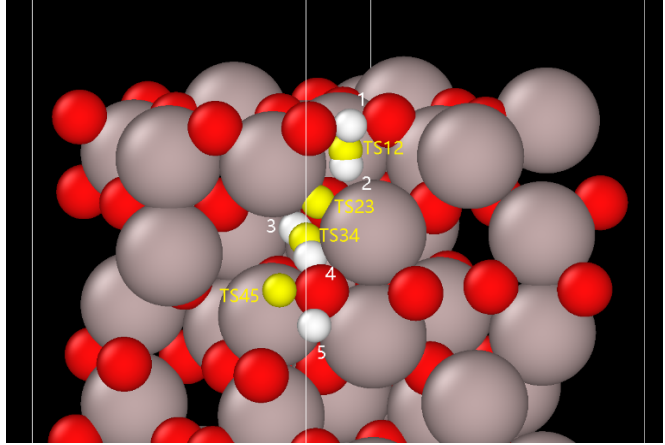

**Fig. 7- Migration path through defect A**

The energy curve and relative energy $E_{rel}$ of the serial sites along the path are demonstrated as "1 H total" in Figure 9 and Table 3 below.

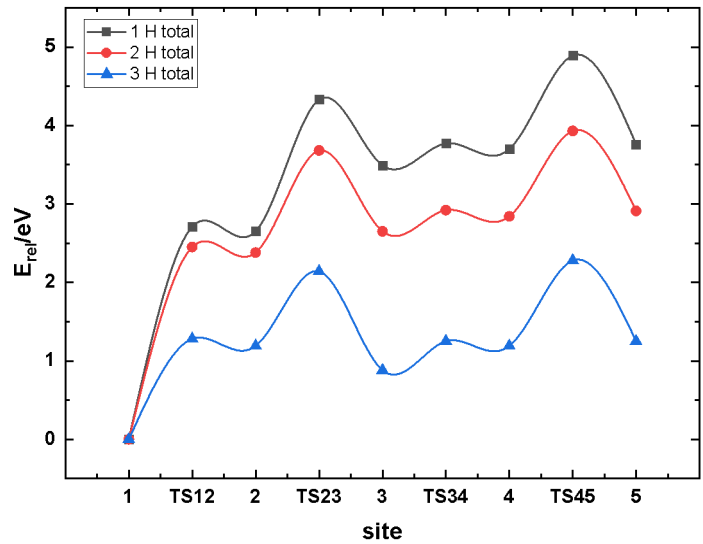

**Fig. 8- Energy curve along the migration path**

**Table 3- Relative energy $E_{rel}$ corresponding to serial sites**

| | site | 1 | TS12 | 2 | TS23 | 3 | TS34 | 4 | TS45 | 5 |
|---|---|---|---|---|---|---|---|---|---|---|
| **$E_{rel}$ /eV** | 1 H total | 0 | 2.71 | 2.65 | 4.33 | 3.49 | 3.77 | 3.70 | 4.89 | 3.76 |
| | 2 H total | 0 | 2.45 | 2.38 | 3.68 | 2.65 | 2.92 | 2.84 | 3.93 | 2.91 |
| | 3 H total | 0 | 1.28 | 1.19 | 2.14 | 0.88 | 1.25 | 1.19 | 2.28 | 1.25 |

Throughout the path, the most stable site is site 1 on the outermost layer where is a deep well. The H atom needs to overcome a large barrier, 2.71 eV, reaching site 2, but the reverse process of this step is much easier, with a much smaller barrier of 0.06 eV, the bond length of H-O at site 2 is 1.18 Å, the weak combination causes the low stability of H atom at site 2 and it can hardly stay there. The next step is to rotate on the subsurface O atom from site 2 to site 3, with a high barrier of 1.68 eV, afterwards, a jump step occurs at site 3 to overcome a low barrier of 0.28 eV to reach site 4, continuously overcomes 1.19 eV barrier from site 4 to reach site 5. At site 5, the H atom forming energy is approximately 4.05 eV, which is close to the value in bulk environment, accordingly it can be considered that the diffusion from site 5 deeper into the slab is in bulk environment.

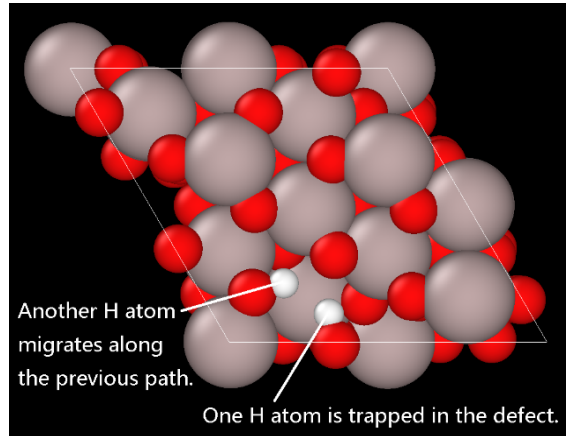

**Fig. 9-2 H total, one is trapped in the defect & the another one migrates**

This result indicates that the point defect in which the surface loses an $Al^{3+}$ in position A will firstly form a deep well, causing a single H atom to fall into the defect area on the outermost layer and stably adsorb on the O atom of outermost layer, thus making it difficult for H atom to permeate through the defect alone. Based on this result, it's necessary to study how the H atoms retained in the defect region and how other H atoms permeate into the slab through the defect with H atom retained. As demonstrated in Figure 10, one H atom is adsorbed on the O atom of the outermost layer in the defect, as described above, another H atom migrates along the same path as previous, permeates into the slab and records each pivotal sites' coordinates, energy, etc. The energy curve and relative energy $E_{rel}$ of the serial sites along the path are demonstrated as "2 H total" in Figure 9 and Table 3. It can be seen from the graph that the retention of the first H atom in the defect area has a certain reducing the ability for defect to bind H atoms. The maximum barrier is still in the jump step from the outermost layer to the subsurface, which is 2.45 eV, compared with the previous curve, this value is reduced by 0.26 eV, besides, the increase and decrease characteristics of the entire energy curve are substantially unchanged.

This means that it's still tough for the second H atom to overcome the barrier of the first jump step to reach site 2, which, like the first H atom, easily bonds strongly to the outermost surface O atom and remains in defect region.

Let two H atoms adsorbed on the outermost layer in defect area, then let the third H atom migrate along the same path as previous, and obtain the energy curve and relative energy $E_{rel}$ of each site along the path as demonstrated by "3 H total" in Figure 9 and Table 3. It can be seen that the adsorption of the third H atom seems to cause the defect region to reach a "saturated" state, at the meantime, the ability of defect to bind H atom is significantly reduced, especially the jump potential from site 1 to the 2, has been greatly reduced, only 1.28 eV, which is already lower than the maximum barrier of 1.53 eV obtained in 3.1.1. Meanwhile, the increase and decrease characteristics of the entire energy curve still almost unchanged from the previous two curves.

It is believed that for surface point defects lacking $Al^{3+}$ in position A, when two H atoms are retained within the defect, the rate at which other H atoms permeate into the slab through the defect will be faster than through perfect surface.

### 3.2.2 An Al3+ lacked at position B on the outermost layer

The diffusion path of H atom is:
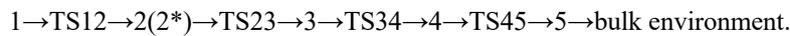
1→TS12→2(2*)→TS23→3→TS34→4→TS45→5→bulk environment.

As demonstrated in Figure 11 below.

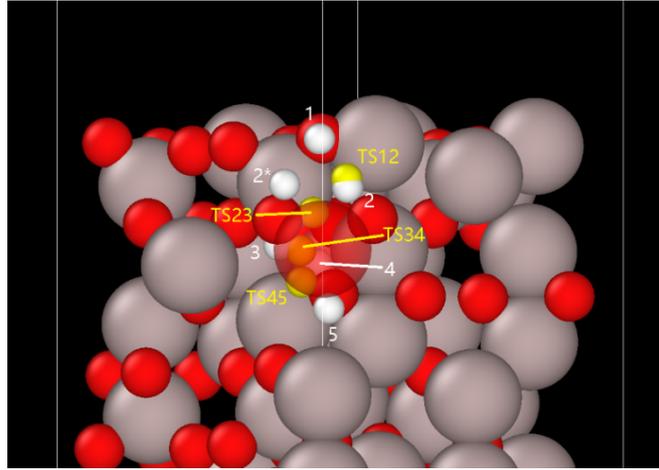

**Fig. 10- Migration path through defect B, site 2 and Site 2* are equivalent sites, with equivalent formation energy and position in the system. This is done to show the path more conveniently.**

The energy curve and relative energy $E_{rel}$ of the serial sites along the path are demonstrated as "1 H total" in Figure 12 and Table 4 below.

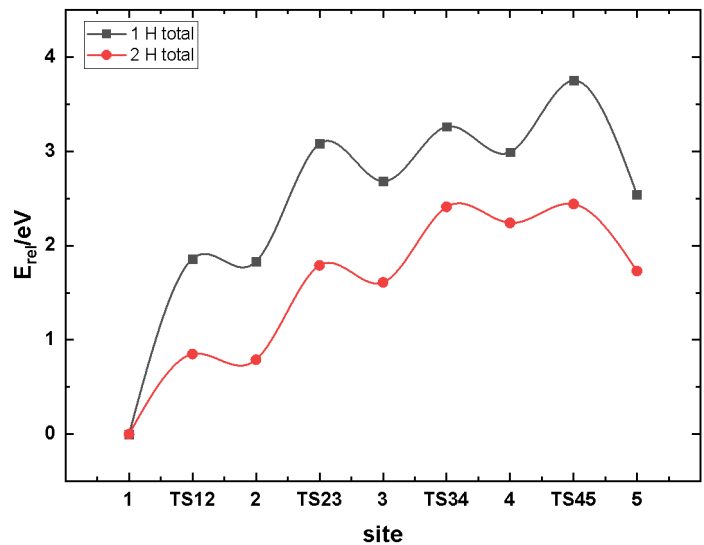

**Fig. 11- Energy curve along the migration path**

**Table 4- Relative energy $E_{rel}$ corresponding to serial sites**

| | site | 1 | TS12 | 2 | TS23 | 3 | TS34 | 4 | TS45 | 5 |
|---|---|---|---|---|---|---|---|---|---|---|
| $E_{rel}$ /eV | 1 H total | 0 | 1.86 | 1.83 | 3.08 | 2.68 | 3.26 | 2.99 | 3.75 | 2.54 |
| | 2 H total | 0 | 0.85 | 0.79 | 1.79 | 1.61 | 2.41 | 2.24 | 2.44 | 1.73 |

In the path of the first H atom permeating through the defect into the slab, the most stable site is still site 1 on the outermost layer. This position is a deep well, where H atom needs to overcome a high barrier of 1.86 eV to arrive at site 2, but the reverse barrier of this step is small, about 0.03 eV, and the bond length of H-O at site 2 is 1.17 Å, meaning the combination is weak and low stability of H atom at site 2 thus it can hardly stay there. These results are extremely similar to the results in 3.2.1. The next step is that the H atom rotates on the subsurface O atom, overcomes a barrier of 1.25 eV from site 2 to site 3, then overcomes 0.58 eV, 0.76 eV two lower barrier arrival sites 4, site 5, successively.

As with 3.2.1, this result indicates that the point defect where the surface loses $Al^{3+}$ in position B will firstly form a deep well, which the potential well is shallower than ion defect A in 3.2.1, but similarly it will cause a single H atom to fall into the defect area and stably adsorb on the O atom of outermost layer, consequently it can hardly permeate through the defect into the material by itself. Based on this result, we let one H atom adsorb on the O atom of the outermost layer in defect area, let another H atom migrate along the previous path, permeate into the slab and record the coordinates and energy of each pivotal site. The energy curve and relative energy $E_{rel}$ of the serial sites along the path are demonstrated as "2 H total" in Figure 12 and Table 4. It can be seen from the graph that the retention of the first H atom in the defect area has a great change in various aspects of the defect. The jump barrier of the new H atom from site 1 to the 2 is only 0.85 eV, furthermore, the maximum barrier is no longer at this step but is located on the rotation step from site 2 to site 3 of the secondary surface layer, which is 1 eV, already lower than the maximum barrier permeates through perfect surface. The increase and decrease characteristics of the entire energy curve are almost unchanged.

Accordingly, it is considered that when a defect region in which $Al^{3+}$ is missing in position B retains one H atom, other H atoms will more easily permeate through defect than through perfect surface.

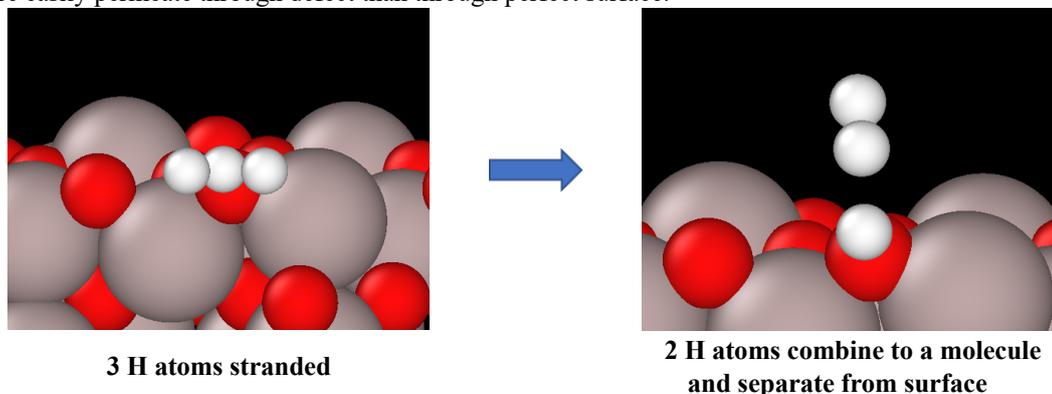

**3 H atoms stranded**      **2 H atoms combine to a molecule and separate from surface**

**Fig. 12-3 H total, not all 3 of them can stably stay in the defect**

When there are 3 H atoms in defect area, the area seems to reach a "supersaturated" state, at this stage, not all of 3 H atoms can be stably adsorbed and retained in the defect. The results of multiple relaxations indicate that there will always be some 2 H atoms likely to detach from the O atoms, close to each other soon to form a $H_2$ molecule, which finally leaves the surface and enter into vacuum.

### 3.2.3 An Al3+ lacked at position C on the outermost layer

The diffusion path of H atom is:
$$1 \rightarrow TS12 \rightarrow 2 \rightarrow TS23 \rightarrow 3 \rightarrow TS34 \rightarrow 4 \rightarrow TS45 \rightarrow 5 \rightarrow \text{bulk environment.}$$
As demonstrated in Figure 14 below.

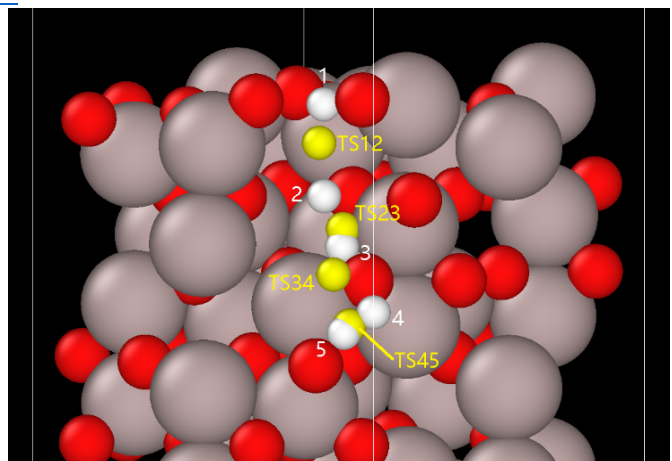

**Fig. 13- Migration path through defect C**

The energy curve and relative energy $E_{rel}$ of the serial sites along the path are demonstrated as "1 H total" in Figure 15 and Table 5 below.

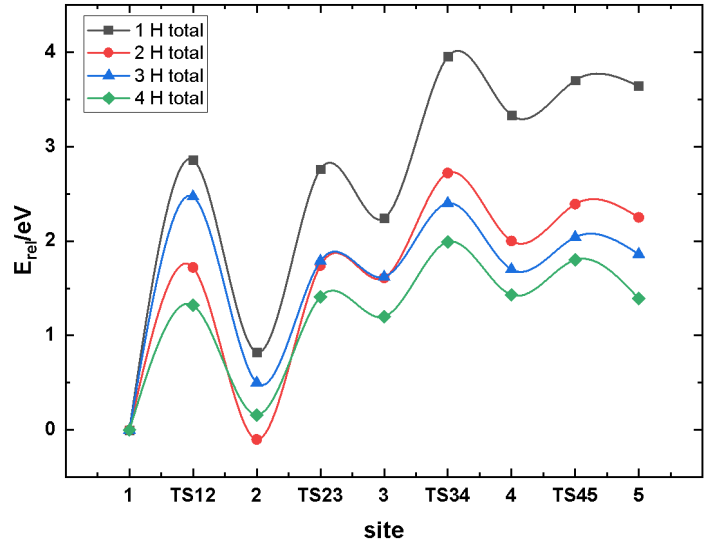

**Fig. 14- Energy curve along the migration path**

**Table 5- Relative energy $E_{rel}$ corresponding to serial sites**

| | site | 1 | TS12 | 2 | TS23 | 3 | TS34 | 4 | TS45 | 5 |
|---|---|---|---|---|---|---|---|---|---|---|
| $E_{rel}$ /eV | 1 H total | 0 | 2.86 | 0.82 | 2.76 | 2.24 | 3.95 | 3.33 | 3.70 | 3.64 |
| | 2 H total | 0 | 1.72 | -0.10 | 1.74 | 1.61 | 2.72 | 2.00 | 2.39 | 2.25 |
| | 3 H total | 0 | 2.47 | 0.50 | 1.79 | 1.62 | 2.40 | 1.70 | 2.04 | 1.86 |
| | 4 H total | 0 | 1.32 | 0.16 | 1.41 | 1.20 | 1.99 | 1.43 | 1.80 | 1.39 |

As can be seen from Figure 14, this defect is significantly different from the previous two defects. The internal volume is larger than the previous two, meanwhile, there are totally 6 O atoms in defect area for H atoms to adsorb. Accordingly, for this defect, a more complicated discussion is required.

For the first H atom's permeation path, the most stable site is still site 1 on the outermost layer. The H atom needs to overcome a very high barrier of 2.86 eV, reaching site 2, besides, the reverse barrier of this step is also large, 2.04 eV. The next step is the jump step from site 2 to site 3, which needs to overcome a high barrier of 1.94 eV, next, rotate around the adsorbed O atom at site 3, overcoming the barrier of 1.71 eV to reach site 4, the below position of the O atom, jump from site 4 to overcome the 0.37 eV barrier to site 5.

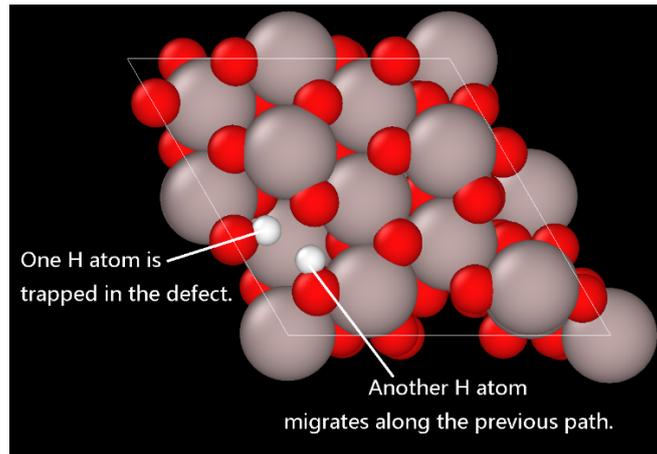

**Fig. 15-2 H total, one is trapped in the defect & the another one migrates**

Similar to the previous two defects, it is exceedingly arduous to overcome the jump barrier of the first step for a single H atom. A point defect in which the surface loses the $Al^{3+}$ in position C forms an extremely deep potential well, causing a single H atom to stably adsorb on the outermost surface O atom, thus it can hardly permeate through the defect into the material by itself.

Therefore, let one H atom adsorb on an O atom of the outermost layer of the defect area and stay there. As in the two previous defects, another H atom migrates along the same path and permeates into the interior of the slab, as shown in Figure 16. Record the coordinates, energy, etc. of each pivotal site. The energy curve and relative energy $E_{rel}$ of the serial sites along the path are demonstrated as "2 H total" in Figure 15 and Table 5.

It can be seen from the graph that the barrier of the first step drops to 1.72 eV, the maximum barrier is no longer at this step, but is in the step jump from the second surface layer site 2 to the deeper layer site 3, about 1.84 eV. The rotation barrier from site 3 to site 4 is 1.11 eV, which is 0.6 eV lower than the value of 1.71 eV of the previous path; the jump barrier from site 4 to site 5 is 0.39 eV, which is basically constant relative to the previous path. The increase and decrease characteristics of the entire energy curve have a certain degree of change compared with the previous one. According to the analysis around Arrhenius formula in introduction, the second coming H atom will no longer be preferentially retained in the outermost layer but will be more likely to remain in the secondary surface layer.

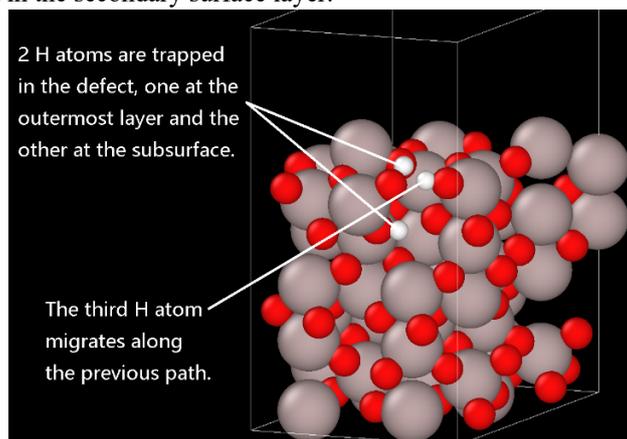

**Fig. 16-3 H total, 2 are trapped in the defect & the third one migrates**

Let the defect area retain two H atoms, one of which is adsorbed on the O atom of the outermost layer (equivalent to site 1), meanwhile, the other is adsorbed on the O atom of subsurface (equivalent to site 2), let the third H atoms migrate along the same path as previous, as shown in Figure 17, besides, the energy curve and relative energy values of the serial sites along the path are demonstrated as "3 H total" in Figure 15 and Table 5.

In this path, the maximum barrier returns to the jump step from site 1 to site 2, which is 2.47 eV, and the jump barrier from site 2 to site 3 is further reduced to 1.29 eV, followed by the rotation step and jump step with barriers of 0.78 eV and 0.34 eV, respectively.

One noteworthy phenomenon is that the barrier of the jump step from site 1 to site 2 has risen sharply from 1.72 eV to 2.47 eV, this is probably because the second H atom stays on the secondary surface layer, the saturation of the subsurface layer of the defect area is increased, meanwhile, the number of H atoms remaining on the outermost layer is still one, consequently the relative potential difference of the subsurface layer to the outermost layer is improved. This means that it's still tough for the third H atom to overcome the barrier of the first jump step to reach site 2, which, like the first H atom, easily bonds strongly to the most surface O atoms and remains in defect area.

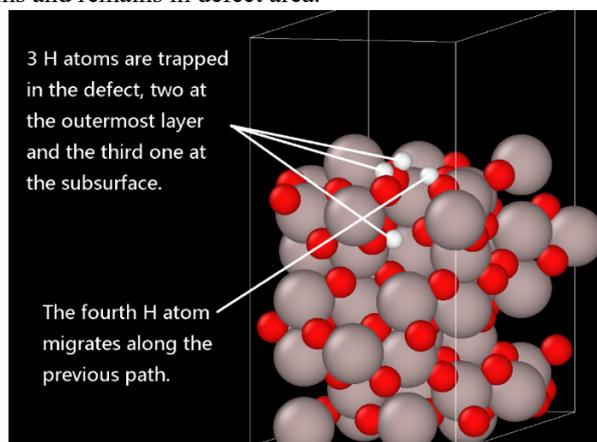

**Fig. 17-4 H total, 3 are trapped in the defect & the fourth one migrates**

Let a total of three H atoms adsorbed in the defect area, two of them are adsorbed on the O atoms of the outermost layer, meanwhile, the third is adsorbed on the O atom of the secondary surface layer, allowing the fourth H atom to migrate along the same path as previous, as shown in Figure 18. The energy curve and relative energy $E_{rel}$ of the serial sites along the path are demonstrated as "4 H total" in Figure 15 and Table 5. It can be seen that the defect approaches saturation and the ability to bind H atoms is significantly reduced at this time, and the jump barrier from site 1 to site 2 is lowered to 1.32 eV, which is already lower than the maximum barrier of 1.53 eV from the perfect surface in 3.1.1. The jump barrier from site 2 to site 3 is 1.25 eV, which is slightly lower than the 1.29 eV in the previous curve. Subsequently rotation barriers from site 3 to site 4 and jump barriers from site 4 to site 5 are 0.79 eV and 0.37 eV, respectively, which are substantially unchanged from the previous curve.

According to this, for the surface point defect lacking $Al^{3+}$ in position C, when three H atoms remain in the defect, two of them are absorbed on the outermost layer, and the third is absorbed on the secondary surface layer, The rate for the fourth H atom permeate into the slab through defects will be faster than through perfect surface.

*3.2.4 Summary of Al3+ point defects*

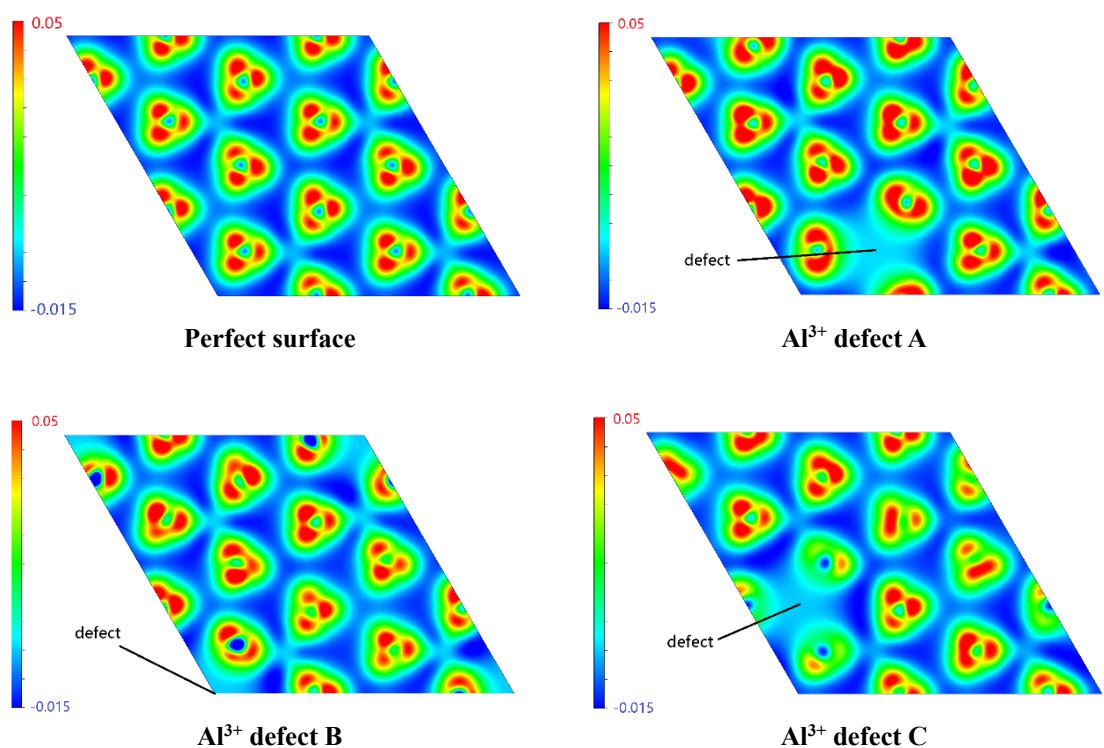

**Perfect surface**  **$Al^{3+}$ defect A**

**$Al^{3+}$ defect B**  **$Al^{3+}$ defect C**

**Fig. 18-**

Figure 19 above are screenshots of deformation charge density of a complete surface and surfaces with three kinds of $Al^{3+}$ point defects. The truncated plane is located at a height of 12 Å from the bottom of the supercells, with the plane vector points to (0001) direction. It can be found from the comparison of the pictures that when an $Al^{3+}$ is missing from the outermost layer, the blue color becomes shallower in defect area, that is, the electronegativity of the defect area is enhanced, a deep well for H atoms is formed, which is the probable reason why the first H atom is easily retained in defect area.

Not all surface $Al^{3+}$ defects possess the same structure, different types (A/B/C) defects, internal volume will be different, thus the amount of O atoms available for H atoms to absorb will also be different, so the amount of H atoms that can be stably accommodated in the defect area will also be different. Newly entering H atoms to defect will weaken the ability of defects to bind H atoms, increase the saturation of the defect, generally reduce the maximum barrier that H atom needs to overcome. When the amount of H atoms in the defect area reaches the upper limit amount that can be accommodated by the defect, the defect looks like reach a state of saturation. At this time, the H atom only needs to overcome a low barrier to migrate, i.e. the permeation into the material through defect will be easier than through perfect surface. If the number of H atoms in the defect continues to increase, the defect looks like reach a "supersaturated" state, the defect region will change from a relative potential well to a relative potential barrier. It will no longer maintain the stable retention of all H atoms, it is likely to appear the phenomenon that some two H atoms of them are detached from the adsorption of O atoms, shortly afterwards, combined with each other to form a $H_2$ molecules, eventually leaving the surface.

### 3.2.5 An O2- lacked at position D on the outermost layer

According to the analysis in 2.2, each O atom located on the outermost layer is equivalent, there should be only one kind of O atoms on the outermost layer.

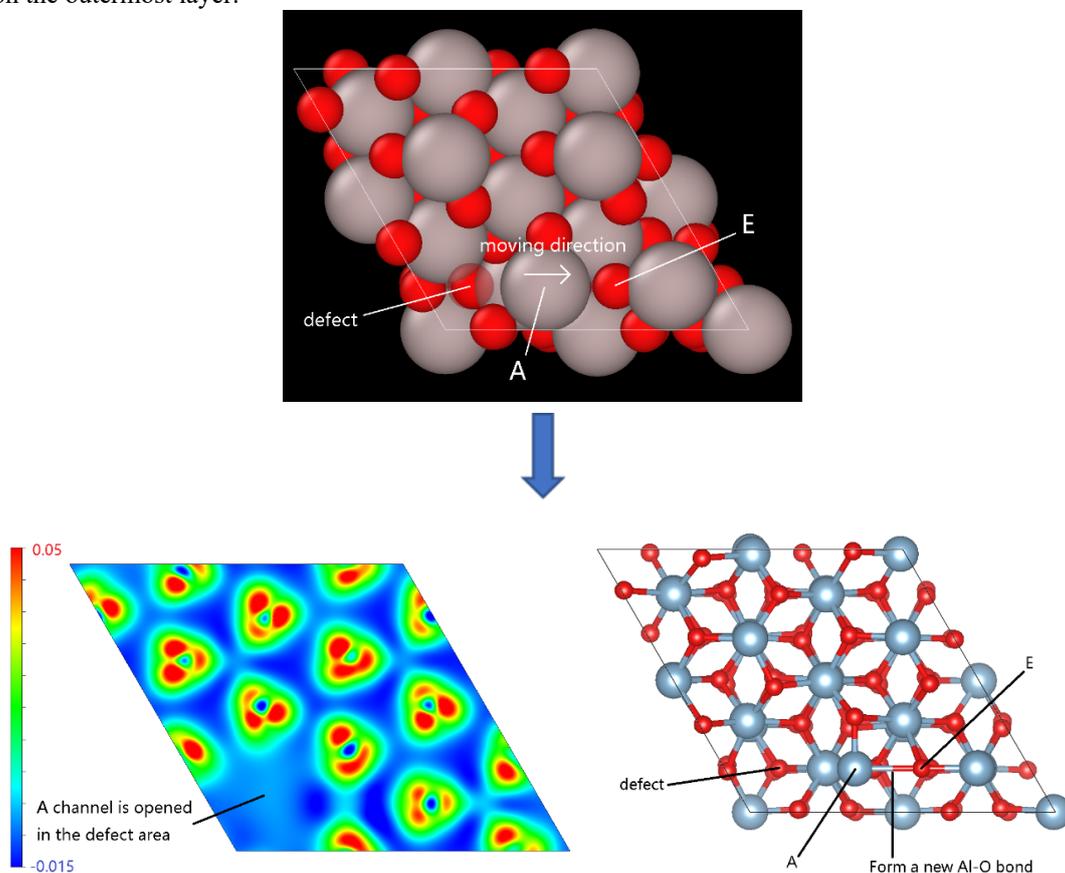

**Fig. 19-**

After relaxation, it is found that when there is an $O^{2-}$ defect in position D of the outermost layer, the Al atom at position A will be dragged by the O atom at position E, then moved by a distance of about 0.76 Å along the direction shown in Figure 20, thereby deviating from the original position and form a new bond with the O atom at position E. This behavior looks like is opening a "channel" for H atoms to permeate, which may be the key reason for the easy entry of H atoms from the outermost layer into the subsurface.

The diffusion path of H atom is:
$$1 \rightarrow TS12 \rightarrow 2 \rightarrow TS23 \rightarrow 3 \rightarrow TS34 \rightarrow 4 \rightarrow TS45 \rightarrow 5 \rightarrow \text{bulk environment}.$$

As demonstrated in Figure 21 below.

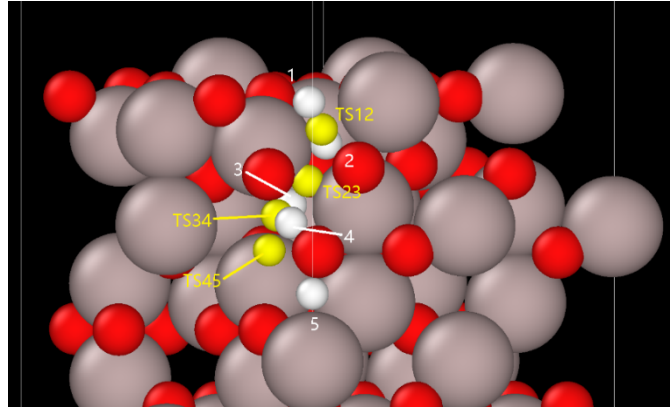

**Fig. 20- Migration path through defect D**

The energy curve and relative energy $E_{rel}$ of the serial sites along the path are demonstrated in Figure 22 and Table 6 below.

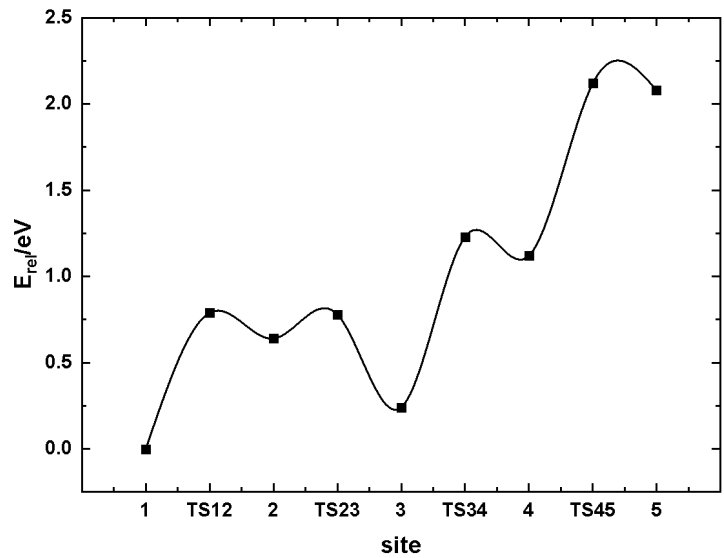

**Fig. 21- Energy curve along the migration path**

**Table 6- Relative energy $E_{rel}$ corresponding to serial sites**

| site | 1 | TS12 | 2 | TS23 | 3 | TS34 | 4 | TS45 | 5 |
|---|---|---|---|---|---|---|---|---|---|
| $E_{rel}$ /eV | 0 | 1.86 | 1.83 | 3.08 | 2.68 | 3.26 | 2.99 | 3.75 | 2.54 |

In this path, the maximum barrier is no longer located at the step from the outermost layer to the subsurface, but rather at the jump step from site 4 to site 5, at 0.99 eV. At site 1, the H atom is located just at the entrance of "channel" where the Al atom A is dragged by the O atom E, so that H atom only needs to overcome a low barrier to move through this channel to site 2, about 0.79 eV. Site 3 is a deep well where the H atom is less energetic and is the most stable site in the entire path except site 1. Then the H atoms follow 3→TS34→4→TS45→5 path, overcoming the two larger barriers, approximately 1 eV, entering the bulk environment. All of these barriers are lower than 1.53eV, the maximum barrier while permeating through perfect surface.

## 4. Conclusion

To find underlying reasons for increase of permeation under irradiation, hydrogen permeation on perfect α-$Al_2O_3$ (0001) surfaces and defected surfaces has been studied by first-principles calculations in this study. For the defected surfaces, three types of Al vacancies and one type of O vacancy are considered. The potential energy pathway of H diffusion is calculated. The results clearly show that the maximum barrier of H permeation will change significantly if vacancy is present on the surface.

If the outermost layer has an $V_{O2-}$, the adjacent Al atom of $V_{O2-}$ will be dragged by a nearby O atom. Consequently, this Al atom will be deviated from its original position, the spacing between these Al atoms increases, leaving a channel for H atom to permeate. As a result, H atoms can easily migrate through the channel into the interior of the material. Both the barrier from outermost layer to subsurface and the maximum barrier of the entire path are significantly reduced.

If $V_{Al3+}$ is present in the outermost layer, the defect region firstly becomes high electronegativity and acts as a deep well for H atom. The barrier for H atom entering the subsurface from the outermost layer is significantly increased, meaning that H atoms will be strongly trapped there and keep stable. However, as more H atoms reach the defect, i.e. the amount of retained H atoms increases, the defect region will gradually approach saturation. Meanwhile, the ability for defect to catch and retain H atoms will decrease. Not all these H atoms can stay in the defect region with high stability. Instead, one of them can migrate through the defect into the material by overcoming a smaller barrier. The maximum amount of H atoms that can be stably accommodated in the defect area depends on the specific type of defect.

In summary, the various point defects on the surface will eventually reduce the barrier of H permeation at different levels. Consequently, H permeation into the material becomes much easier. Point defects on surface are likely to be the underlying reason that causes a large decrease in PRF when α-$Al_2O_3$ is under irradiation. For future investigation, the effects of surface defects on the adsorption and dissociation of hydrogen molecule need to be studied. Together with this study, a complete understanding of how radiation-induced point defects affect H permeation could probably be acheived.

## Acknowledgements

The authors would like to acknowledge the support from the National Natural Science Foundation of China (Grant No. 11905071).